\documentclass[twocolumn]{article}
\let\ifapj\iffalse
\let\ifarxiv\iftrue
\let\iflocal\iffalse
\ifapj\usepackage[strict]{patch-apj}\NewPageAfterKeywords\fi
\usepackage{etoolbox,ifxetex,ifluatex}
\ifxetex\else\ifluatex\else\usepackage[utf8]{inputenc}\fi\fi
\ifluatex\AtEndPreamble{\hypersetup{pdfencoding=auto}}\fi
\usepackage{siunitx}
\ifapj
  \usepackage{acro,csquotes,hyperref}
  \let\textacsize\relax
\fi
\ifboolexpr{bool{arxiv} or bool{local}}{
  \usepackage{font-termes,basic,physics,astro}
  \usepackage{draft}
  \usepackage[mode=biblatex, hyperref]{apjbib}
  \ExecuteBibliographyOptions{embedlinks}
  \addbibresource{tde.bib}
}{}
\DeclareAcronym{TDE}{
  short           =TDE,
  long            =tidal disruption event}
\DeclareAcronym{AGN}{
  short           =AGN,
  long            =active galactic nucleus,
  short-indefinite=an,
  long-indefinite =an,
  long-plural-form=active galactic nuclei}
\DeclareAcronym{UV}{
  short           =UV,
  long            =ultraviolet,
  long-indefinite =an}
\newcommand*\DeclareSubscript[3]{\newcommand*#1{#2_\su{#3}}}
\DeclareSubscript\smc{\dot M}s
\DeclareSubscript\dmc{\dot M}d
\DeclareSubscript\amc{\dot M}a
\newcommand*\mcr{\smc/\dmc}
\DeclareSubscript\Mh Mh
\DeclareSubscript\Ms M\star
\DeclareSubscript\rs r\star
\DeclareSubscript\Rt{\mathcal R}t
\DeclareSubscript\rp rp
\DeclareSubscript\rg rg
\DeclareSubscript\LE LE
\newcommand*\mH{m_\element H}
\newcommand*\me{m_\particle e}
\DeclareSubscript\rhof\rho f
\DeclareSubscript\vf{\vec v}f
\DeclareSubscript\Fd Fd
\newcommand*\vfz{v_{\su f,z}}
\DeclareSubscript\Eg Eg
\DeclareSubscript\Er Er
\newcommand*\Einfty{E_\infty}
\DeclareSubscript\Hsh H{sh}
\DeclareSubscript\tdep t{dep}
\DeclareSubscript\tsca t{sca}
\DeclareSubscript\tesc t{esc}
\DeclareSubscript\vsh v{sh}
\DeclareSubscript\Heq H{eq}
\DeclareSubscript\cs cs
\newcommand*\nelc{n_\particle e}
\DeclareSubscript\tauT\tau T
\DeclareSubscript\Sigmash\Sigma{sh}
\DeclareSubscript\Tg Tg
\DeclareSubscript\Td Td
\DeclareSubscript\FC FC
\DeclareSubscript\ThetaC\Theta C
\DeclareSubscript\Teff T{ef{}f}
\newcommand*\NH{N_\element H}
\newcommand*\xray{X\nobreakdash-ray}
\newcommand*\xrays{X\nobreakdash-rays}
\newcommand*\gammarays{$\gamma$\nobreakdash-rays}
\newcommand*\spandsp{ and }
\ifboolexpr{bool{arxiv} or bool{local}}{
  \usepackage{tikz}
  \usetikzlibrary{
    decorations.pathmorphing, decorations.pathreplacing, shapes.arrows}
  \tikzset{
    photon/.pic={
      \draw[->, decorate, decoration={snake, segment length=1ex,
      amplitude=.3ex, pre length=.5ex-1pt, post length=.5ex}] (0,0) -- (3ex,0);
    },
    pics/triad out/.style n args={3}{code={
      \draw circle (.4ex);
      \fill circle (\pgflinewidth*1.5);
      \draw[->] (.4ex,0) -- (2ex,0);
      \draw[->] (0,.4ex) -- (0,2ex);
      \node at (3ex,0) {$\vphantom x\smash{#1}$};
      \node at (0,3ex) {$\vphantom x\smash{#2}$};
      \node at (-135:1.6ex) {$\vphantom x\smash{#3}$};
    }},
    pics/triad in/.style n args={3}{code={
      \draw circle (.4ex);
      \draw (45:.4ex) -- (-135:.4ex) (-45:.4ex) -- (135:.4ex);
      \draw[->] (.4ex,0) -- (2ex,0);
      \draw[->] (0,.4ex) -- (0,2ex);
      \node at (3ex,0) {$\vphantom x\smash{#1}$};
      \node at (0,3ex) {$\vphantom x\smash{#2}$};
      \node at (-135:1.6ex) {$\vphantom x\smash{#3}$};
    }}}
}{}
\ifapj\let\edit\relax\else\usepackage{xcolor}\fi
\expandafter\def\csname editcolor1\endcsname{blue!60!cyan}
\newcommand*\edit[2]{\textcolor{\csname editcolor#1\endcsname}{#2}}
\begin{document}

\title{High-energy emission from tidal disruption events in active galactic
nuclei}

\ifapj
  \author[0000-0001-5949-6109]{Chi-Ho Chan}
  \affiliation{Racah Institute of Physics, Hebrew University of Jerusalem,
  Jerusalem 91904, Israel}
  \affiliation{School of Physics and Astronomy, Tel Aviv University,
  Tel Aviv 69978, Israel}
  \author[0000-0002-7964-5420]{Tsvi Piran}
  \affiliation{Racah Institute of Physics, Hebrew University of Jerusalem,
  Jerusalem 91904, Israel}
  \author[0000-0002-2995-7717]{Julian~H. Krolik}
  \affiliation{Department of Physics and Astronomy, Johns Hopkins University,
  Baltimore, MD 21218, USA}
\fi

\ifboolexpr{bool{arxiv} or bool{local}}{
  \author[1,2]{Chi-Ho Chan}
  \author[1]{Tsvi Piran}
  \author[3]{Julian~H. Krolik}
  \affil[1]{Racah Institute of Physics, Hebrew University of Jerusalem,
  Jerusalem 91904, Israel}
  \affil[2]{School of Physics and Astronomy, Tel Aviv University,
  Tel Aviv 69978, Israel}
  \affil[3]{Department of Physics and Astronomy, Johns Hopkins University,
  Baltimore, MD 21218, USA}
}{}

\date{April 29, 2021}

\shorttitle{High-energy emission from TDEs in AGNs}
\shortauthors{Chan et al.}
\pdftitle{High-energy emission from tidal disruption events in active galactic
nuclei}
\pdfauthors{Chi-Ho Chan, Tsvi Piran, Julian Krolik}

\begin{abstract}
\Acp{TDE} taking place in \acp{AGN} are different from ordinary \acp{TDE}. In
these events, the returning tidal debris stream drills through the pre-existing
\ac{AGN} accretion disk near the stream pericenter, destroying the inner disk
in the process, and then intersects with the disk a second time at radii
ranging from a few to hundreds of times the pericenter distance. The debris
dynamics of such \acp{TDE}, and hence their appearance, are distinct from
ordinary \acp{TDE}. Here we explore the observational signatures of this
\textquote{second impact} of the stream with the disk. Strong shocks form as
the dilute stream is stopped by the denser disk. Compton cooling of the shocked
material produces hard \xrays, even soft \gammarays, with most of the energy
emitted between \SI{\sim10}{\kilo\electronvolt} and \SI{1}{\mega\electronvolt}.
The luminosity follows the mass-return rate, peaking between
\SIrange[range-phrase=\spandsp]{\sim e42}{e44}{\erg\per\second}. The \xray{}
hardness and the smoothness of the light curve provide possible means for
distinguishing the second impact from ordinary \ac{AGN} flares, which exhibit
softer spectra and more irregular light curves.
\end{abstract}
\acresetall

\section{Introduction}

Supermassive black holes at galactic centers can generate copious amounts of
radiation. When a black hole is fed a steady diet of gas from galactic scales,
the gas forms an accretion disk around the black hole, and the black hole
manifests itself as \iac{AGN} that shines across the entire electromagnetic
spectrum. The luminosity of \iac{AGN} varies over time, exhibiting variations
from mundane, \SI{\sim10}{\percent} fluctuations to more dramatic flares.

Once in a while, a star in the nuclear cluster of the host galaxy may be
scattered onto an orbit grazing the black hole. \Iac{TDE} takes place when the
tidal gravity of the black hole overwhelms the self-gravity of the star and
breaks the star apart \citep{1975Natur.254..295H}. Energy redistribution during
the disruption leaves roughly half of the star gravitationally bound to the
black hole while unbinding the other half \citep[e.g.,][]{1988Natur.333..523R}.
If the black hole is not surrounded by a disk, the bound debris returns to the
neighborhood of the black hole as an elongated stream, shocks against itself,
and settles into an accretion flow that emits primarily in the optical/\ac{UV}.
The unbound debris produces synchrotron radiation through its interaction with
the circumnuclear medium \citep{2016ApJ...827..127K, 2019MNRAS.487.4083Y}.

If the disruption takes place in \iac{AGN} with a pre-existing disk, the system
initially evolves almost as if the disk were absent. Because one large-angle
two-body scattering in the nuclear star cluster suffices to reduce a star's
pericenter to within the tidal radius, a typical star does not pass through the
small-scale disk multiple times before disruption \citep{2020ApJ...889...94M}.
The mutual influence between the star and the disk in a single transit is
extremely limited: the internal structure of the star is barely altered
\citep{1991MNRAS.250..505S, 2005ApJ...619...30M}; the compact star scarcely
opens a gap in the disk \citep{2021arXiv210208135S}; and hydrodynamic drag
hardly changes the stellar orbit \citep{1991MNRAS.250..505S,
1995MNRAS.275..628R, 2005ApJ...619...30M, 2012ApJ...758...51J}. This means both
the star and the disk are fully intact before the disruption, and the star's
trajectory is independent of the disk's orientation. The disruption itself is
also unaffected by the disk, due to the much higher density of the star and the
debris.

The story after the debris stream falls back to pericenter is strongly modified
because the debris at this point is much more dilute. The most vigorous
interaction between stellar and disk material happens when the stream makes
first contact with the disk. Stream fallback and disk rotation carry comparable
mass currents
\citetext{\citealp{2019ApJ...881..113C}\multicitedelim\citealp[see
also][]{1994ApJ...422..508K}}, and the stream is also geometrically extended;
both properties are critical for stream impact to significantly reshape the
disk. At the same time, the presence of the disk impedes stream
self-intersection, so the observational signatures of \iac{TDE} in \iac{AGN}
may not resemble one in vacuum at all.

In addition to their possibly different appearance, the challenge of finding
\acp{TDE} in \acp{AGN} is compounded by the fact that \acp{AGN} exhibit
frequent flares. Strategies for separating out \acp{TDE} usually depend on
light-curve timescales and optical colors \citep[e.g.,][]{2021ApJ...908....4V},
but the nature of any one event is often disputed \citep{2015JHEAp...7..148K,
2017NatAs...1..865K, 2018ApJ...852...37A, 2019NatAs...3..242T,
2020arXiv201008554F}. This underlines the importance of understanding how
\acp{TDE} in \acp{AGN} produce their light.

We were the first to explore the hydrodynamic and radiative properties of
\acp{TDE} in \acp{AGN} \citep{2019ApJ...881..113C, 2020ApJ...903...17C}. We
call the pericentric collision of the returning stream with the disk the
\textit{first impact}. An important quantity to consider is the disk mass
interior to the impact point. For example, for a black hole of mass
$\Mh=\SI{3e6}{\solarmass}$ and a Sun-like star, the physical tidal radius
$\Rt$, or the maximum pericenter for total disruption, is
$\mathrelp\sim12\,\rg$ \citep{2020ApJ...904...98R}, where $\rg\eqdef G\Mh/c^2$
is the gravitational radius. A \citet{1973A&A....24..337S} disk accreting at
0.01 times Eddington has a meager \SI{\sim e-4}{\solarmass} inside this radius.
It follows immediately that the first impact can significantly reshape the
inner disk.

This expectation is confirmed by the hydrodynamics simulations in
\citet{2019ApJ...881..113C}. Shocks emanating from the first impact dissipate
energy and remove angular momentum so vigorously that they vacate the inner
disk within tens of days. The dissipated energy keeps the bolometric luminosity
constant at Eddington levels for a similar amount of time
\citep{2020ApJ...903...17C}. Unfortunately, the complicated hydrodynamic and
radiative environment of the inner disk prevents us from estimating
first-impact spectra without detailed simulations and radiative-transfer
calculations.

\begin{figure}
\includegraphics{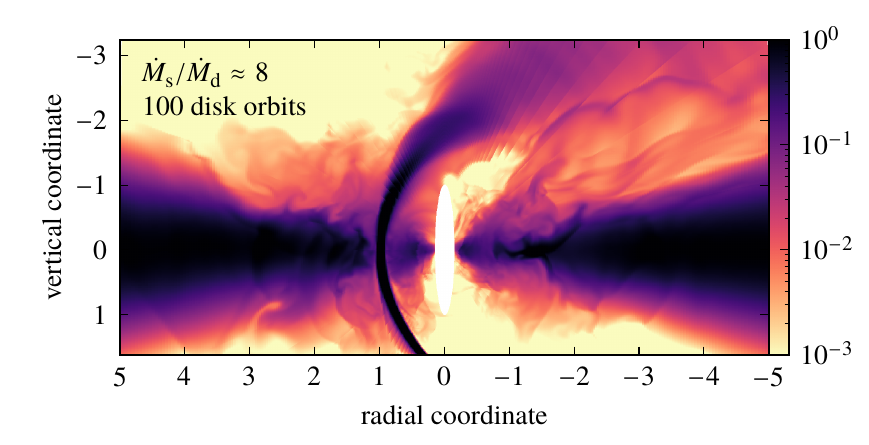}\vskip1ex
\ifapj\includegraphics{stream2}\else
\begin{tikzpicture}[x=.25in, y=.35in, scale=-1, black,
  nodes={font=\footnotesize}]
\pgfmathsetmacro\fxl{8.7}
\pgfmathsetmacro\fxr{4.8}
\pgfmathsetmacro\sru{5}
\pgfmathsetmacro\szl{-3.2}
\pgfmathsetmacro\szu{1.6}
\pgfmathsetmacro\rpu{0.98}
\pgfmathsetmacro\rpv{1.02}
\pgfmathsetmacro\rpw{0.98}
\pgfmathsetmacro\rpx{1.02}
\pgfmathsetmacro\rpy{1.02}
\pgfmathsetmacro\rau{50}
\pgfmathsetmacro\rav{50}
\pgfmathsetmacro\raw{4}
\pgfmathsetmacro\rax{12}
\pgfmathsetmacro\ray{1000}
\pgfmathsetmacro\puv{90}
\pgfmathsetmacro\pwxy{3}
\pgfmathsetmacro\au{(\rpu+\rau)/2}
\pgfmathsetmacro\av{(\rpv+\rav)/2}
\pgfmathsetmacro\aw{(\rpw+\raw)/2}
\pgfmathsetmacro\ax{(\rpx+\rax)/2}
\pgfmathsetmacro\ay{(\rpy+\ray)/2}
\pgfmathsetmacro\bu{sqrt(\rpu*\rau)}
\pgfmathsetmacro\bv{sqrt(\rpv*\rav)}
\pgfmathsetmacro\bw{sqrt(\rpw*\raw)}
\pgfmathsetmacro\bx{sqrt(\rpx*\rax)}
\pgfmathsetmacro\by{sqrt(\rpy*\ray)}
\pgfmathsetmacro\Eu{atan(sqrt(\rpu/\rau)*tan( \puv      /2))*2}
\pgfmathsetmacro\Ev{atan(sqrt(\rpv/\rav)*tan( \puv      /2))*2}
\pgfmathsetmacro\Ew{atan(sqrt(\rpw/\raw)*tan((\pwxy-180)/2))*2}
\pgfmathsetmacro\Ex{atan(sqrt(\rpx/\rax)*tan((\pwxy-180)/2))*2}
\pgfmathsetmacro\Ey{atan(sqrt(\rpy/\ray)*tan((\pwxy-180)/2))*2}
\pgfmathsetmacro\Es{atan(sqrt(\rpv/\rav)*tan( \pwxy     /2))*2}
\pgfmathsetmacro\Et{atan(sqrt(\rpx/\rax)*tan(-\pwxy     /2))*2}
\pgfmathsetmacro\trl{\raw-0.5}
\pgfmathsetmacro\tru{\fxl+1}
\pgfmathsetmacro\tzl{-\fxl*tan(\pwxy)-0.15}
\pgfmathsetmacro\tzu{0}
\pgfmathsetmacro\urc{8}
\pgfmathsetmacro\urd{.2ex/\csname pgf@xx\endcsname}
\fill[black] circle (.5ex);
\begin{scope}[black!20]
\fill
  (-\fxl,{\fxl*tan(\pwxy)})
    --  (180-\pwxy:1)
    arc (180-\pwxy:180+\pwxy:1)
    --  (-\fxl,{-\fxl*tan(\pwxy)})
    coordinate (A2);
\fill
  (\fxr,{\fxr*tan(\pwxy)})
    --  ([shift={(\rpv-\av,0)}] \Es:{\av} and \bv)
    arc (\Es:  0:{\av} and \bv)
    arc (  0:\Et:{\ax} and \bx)
    --  (\fxr,{-\fxr*tan(\pwxy)});
\end{scope}
\begin{scope}[overlay]
\clip (-\fxl,-10) rectangle (\fxr,10);
\fill[blue!50!cyan!60]
  ([shift={(\rpu-\au,0)}] \Eu:{\au} and \bu)
    arc (\Eu:  0:{\au} and \bu)
    arc (  0:\Ew:{\aw} and \bw)
    coordinate (A1)
    coordinate [pos=0.75] (C1) --
  ([shift={(\rpx-\ax,0)}] \Ex:{\ax} and \bx)
    arc (\Ex:  0:{\ax} and \bx)
    arc (  0:\Ev:{\av} and \bv) -- cycle;
\fill[green!60!black!40]
  ([shift={(\rpx-\ax,0)}] \Ex:{\ax} and \bx)
    arc (\Ex:  0:{\ax} and \bx)
    arc (  0:\Ey:{\ay} and \by) -- cycle;
\draw [densely dotted] (-\sru,\szl) rectangle ( \sru,\szu);
\draw [densely dashed] (-\trl,\tzl) rectangle (-\tru,\tzu);
\draw [densely dashed, dash phase=-{mod((\tzu-\tzl)*.35*72.27+2,5)/2}pt]
  (-\urc-\urd,\tzl) -- (-\urc-\urd,\tzu)
  (-\urc+\urd,\tzl) -- (-\urc+\urd,\tzu);
\end{scope}
\coordinate (C2) at
  (-\fxl,{-2*sqrt(\rpy*\ray)/(\rpy+\ray)*sqrt((\rpy+\fxl)*(\ray-\fxl))});
\node[below=1ex, align=center] {black\\hole};
\node at (\fxr,0) [right] {outer disk};
\draw[<-, out=90, in=0] (-0.5,0) to +(135:8ex)
  node [right, align=center] {cleared\\inner disk};
\node at ([shift={(\rpu-\au,0)}] \Eu:{\au} and \bu)
  [below, align=center] {unperturbed\\stream};
\draw[<-, out=0, in=90] (1,0) +(.5ex,0) to +(-45:6ex)
  node [above, align=center] {first\\impact};
\path (A1) -- (A2)
  coordinate [pos=0.025] (B1)
  coordinate [pos=0.975] (B2);
\path (C1) -- (C2)
  node [pos=0.15, align=center, shift={(-1ex,1ex)}] {bound\\material}
  node [pos=0.6,  align=center] {unbound\\material};
\node at (\fxr,\szl)
  [above right, align=left] {domain of\\hydrodynamics simulation};
\node at (-\fxl,\tzl)
  [above left, align=right] {domain of\\second-impact calculations};
\draw[<-] (-\urc,\tzu) ++(0,.5ex) -- +(0,3ex)
  node [below, align=center] {focus of\\\cref{fig:shocked}};
\draw[<->, yshift=-.75ex] (0,-1ex) -- (1,-1ex)
  node [midway, above] {$\rp$};
\pic[shift={(4ex,-3ex)}, rotate=180] at (current bounding box.north west)
  {triad in=xzy};
\end{tikzpicture}
\fi
\caption{Interaction of \iac{TDE} stream with \iac{AGN} disk. \textit{Top
panel:} Density slice of one of the hydrodynamics simulations this work is
based upon, set up according to \citet{2019ApJ...881..113C}. The slice is
rotated upside-down in anticipation of \cref{sec:overall}. The white ellipse
around the origin is due to the numerical cutout, whose purpose is to exclude
the cylindrical axis from the simulation domain. The stream penetrates the disk
at the first impact. Shocks excited there nearly evacuate the disk interior to
the impact point. Subjected to tidal compression during its pericenter passage,
internal pressure forces widen the stream on exit from the disk. \textit{Bottom
panel:} Schematic diagram of the framework employed to calculate the properties
of the second impact. Pressure forces internal to the stream widen its energy
distribution. Most of the stream material is gravitationally unbound, and the
rest is dispersed over a large swath of the outer disk. The much lower density
of the stream at this second impact means it can be completely absorbed by the
disk. The skinny dashed box picks out an arbitrary annulus of the disk;
\cref{fig:shocked} shows the detailed structure of the region inside the box.}
\label{fig:stream}
\end{figure}

There is, however, another way by which \acp{TDE} in \acp{AGN} can glow. Most
streams have enough inertia to punch through the disk, all the more so after
shocks have gotten rid of the inner disk. As illustrated in \cref{fig:stream},
the stream, tidally compressed during its pericenter passage and ruffled by the
first impact, spews out fan-shaped from the other side. As we shall see later,
stream material can be sprayed as far as \num{\sim100} times the pericenter
distance, and when it finally meets the disk again, it is much more dilute than
the disk it lands on. The relation between stream mass and disk mass is also
very different: for the example above, the disk within $\mathrelp\sim100\,\Rt$
contains \SI{\sim17}{\solarmass}. Even though the disk does not carry much more
mass than the stream, it is still much denser, mostly because all that mass is
concentrated within a small scale height. Because of the density disparity, the
stream is likely absorbed in its entirety by the disk without altering the disk
to any appreciable extent.

The energy dissipated at this \textit{second impact} powers another flare, on
top of that from the first impact. Different circumstances suggest different
observational signatures; specifically, the simpler geometry of the second
impact allows us to more easily model the cooling process and compute the
spectrum.

We begin by accentuating in \cref{sec:comparison} the dissimilar nature of
\acp{TDE} in vacuum and \acp{TDE} in \acp{AGN}. Our investigation of the second
impact begins in \cref{sec:parameters} with a survey of system parameters. The
dynamical and energetic aspects of the second impact are addressed in
\cref{sec:second impact}. The core of our exposition is \cref{sec:cooling},
where we translate the energy dissipation rates from the previous section to
spectra using a simple Compton-cooling model. A discussion of the results is
found in \cref{sec:discussion}, followed by our conclusions in
\cref{sec:conclusions}.

\section{TDEs in AGNs as a distinct population from TDEs in vacuum}
\label{sec:comparison}

A primary point of contention about \acp{TDE} in vacuum is whether shocks can
transform the eccentric stream to a more circular accretion flow and, if so,
how. In the classical picture, general-relativistic apsidal precession makes
the stream self-intersect near pericenter; the resulting strong shocks
dissipate energy briskly, forcing the stream to circularize and settle into a
compact accretion disk \citep[e.g.,][]{1988Natur.333..523R}. The radiation from
this disk might be reprocessed by surrounding optically thick matter
\citetext{\citealp[e.g.,][]{1997ApJ...489..573L, 2016MNRAS.461..948M,
2020MNRAS.492..686L}; but see \citealp{2021MNRAS.502.3385M}}. However, if the
disruption takes place at $\mathrelp\gtrsim10\,\rg$, precession leads instead
to shocks that are closer to apocenter \citep{2015ApJ...804...85S,
2015ApJ...812L..39D}. These apocentric shocks are the result of the stream
interacting with the complex, eccentric accretion flow created by
earlier-returning tidal debris. They are too weak to efficiently circularize
the stream, but they can generate optical/\ac{UV} light
\citep{2015ApJ...806..164P}.

The picture is fundamentally different for \acp{TDE} in \acp{AGN}. The
obstruction posed by the disk means the stream cannot in general
self-intersect; specifically, the first impact prevents rapid circularization,
while the second impact precludes apocentric shocks. Mechanisms discussed in
the context of \acp{TDE} in vacuum, including reprocessing and apocentric
shocks as the origin of optical/\ac{UV} emission, are therefore wholly
inapplicable to \acp{TDE} in \acp{AGN}.

The physics of \acp{TDE} in \acp{AGN} must be studied in its own right. The
most energetic stream--disk interaction comes about when the stream first
returns to pericenter, at the first impact. Shocks excited by the impact
accelerate inflow in the inner disk, which leads to vigorous energy
dissipation, often super-Eddington in power \citep{2019ApJ...881..113C}. The
bolometric luminosity depends on a balance between energy dissipation,
radiation diffusion, and photon trapping \citep{2020ApJ...903...17C}. These
mechanisms operate in qualitatively the same way no matter how close the stream
approaches to the black hole, or how much apsidal precession it suffers on its
pericentric flyby. The second impact, the focus of this article, happens when
the stream is arrested near its apocenter as it flies over the disk. As
expanded upon in \cref{sec:spectra}, the second-impact shocks likely emit
primarily hard \xrays{} to soft \gammarays.

Even the energy reservoir is different for the two kinds of \acp{TDE}. For
\acp{TDE} in vacuum, the energy radiated by the bound debris must ultimately be
sourced from its orbital energy. By contrast, \acp{TDE} in \acp{AGN} can tap
into the orbital energy of the disk as well. In most \acp{TDE} of the latter
kind, the returning debris plays the role of a catalyst, causing orbital energy
to be liberated at much higher rates than in the unperturbed disk. This
statement is true for the first impact, whose near-Eddington luminosity results
from the speedy infall of the inner disk \citep{2019ApJ...881..113C,
2020ApJ...903...17C}. It is also true for the second impact: we shall see in
\cref{sec:dynamics} that stream material dissipates the kinetic energy of the
part of the disk it lands on, at a rate proportional to its mass current.

Thus, \textit{\acp{TDE} in vacuum and \acp{TDE} in \acp{AGN} have entirely
distinct appearances.} Searches tailored to one type may be insensitive to the
other. To find \acp{TDE} in \acp{AGN}, a completely new set of distinguishing
features must be constructed for this novel class of transients. In addition,
there is a second difficulty: winnowing away the cases in which
\textquote{flares} are merely intrinsic \ac{AGN} variability.

\section{Parameters}
\label{sec:parameters}

We consider the case where the pericenter distance $\rp$ of the debris stream
equals the physical tidal radius $\Rt$, the maximum pericenter for complete
disruption. This radius is a function of the black-hole mass $\Mh$ and the
stellar mass $\Ms$. Both $\Rt$ and the peak rate $\smc$ at which bound tidal
debris returns to pericenter are given by the following expressions, derived
from general-relativistic hydrodynamics simulations of the disruption of stars
with realistic internal structure \citep{2020ApJ...904...98R}:
\begin{align}
\Rt &\approx \SI{1.0e13}{\centi\meter}\times\Psi
  \biggl(\frac\Mh{\SI{3e6}{\solarmass}}\biggr)^{1/3}
  \biggl(\frac\Ms{\si{\solarmass}}\biggr)^{-1/3}
  \biggl(\frac\rs{\si{\solarradius}}\biggr), \\
\smc &\approx \SI{1.7}{\solarmass\per\year}\times\Xi^{3/2}
  \biggl(\frac\Mh{\SI{3e6}{\solarmass}}\biggr)^{-1/2}
  \biggl(\frac\Ms{\si{\solarmass}}\biggr)^2
  \biggl(\frac\rs{\si{\solarradius}}\biggr)^{-3/2}.
\end{align}
Unlike \citet{2020ApJ...904...98R}, here we use \SI{3e6}{\solarmass} as our
fiducial $\Mh$. In these equations,
\begin{equation}
\rs\approx\SI{0.93}{\solarradius}\times
  \biggl(\frac\Ms{\si{\solarmass}}\biggr)^{0.88}
\end{equation}
is the stellar radius \citep{2020ApJ...904...99R}, and the dimensionless
factors
\begin{align}
\Psi(\Mh,\Ms) &\eqdef
  \{0.80+0.45\,[\Mh/(\SI{3e6}{\solarmass})]^{0.50}\}\times{} \nonumber\\
&\cont \frac{1.47+\exp((\Ms/\si{\solarmass}-0.67)/0.14)}
  {1+2.34\exp((\Ms/\si{\solarmass}-0.67)/0.14)}, \\
\Xi(\Mh,\Ms) &\eqdef
  \{1.27-0.21\,[\Mh/(\SI{3e6}{\solarmass})]^{0.24}\}\times{} \nonumber\\
&\cont \frac{0.62+\exp((\Ms/\si{\solarmass}-0.67)/0.21)}
  {1+0.55\exp((\Ms/\si{\solarmass}-0.67)/0.21)}
\end{align}
refine earlier order-of-magnitude estimates
\citep[e.g.,][]{1988Natur.333..523R}. For $\Mh=\SI{3e6}{\solarmass}$ and
$\Ms=\si{\solarmass}$, the revised $\rp$ is $\mathrelp\approx12\,\rg$, roughly
half the value used in \citet{2019ApJ...881..113C, 2020ApJ...903...17C}. This
changes the dependence of timescales and stream mass current on \ac{TDE}
parameters.

In our previous work, we investigated the interaction of the stream with the
disk at the first impact using a suite of Newtonian hydrodynamics simulations
\citep{2019ApJ...881..113C}. In those simulations, a disk was set up along the
midplane, and a stream on a parabolic orbit was injected from above in such a
way that it made perpendicular contact with the disk exactly when it reached
pericenter. Because the simulations lasted only a fraction of the mass-return
time, the stream mass current was kept constant at $\smc$. The same simulation
setup is employed to study the second impact.

The chief parameter characterizing both impacts is $\mcr$, where $\dmc$ is the
mass current of the unperturbed disk rotating under the stream footprint at the
first impact. Assuming a \citet{1973A&A....24..337S} disk, we found that $\mcr$
has a minimum, and $\mcr\gtrsim1$ for reasonable \ac{TDE} and \ac{AGN}
parameters \citep{2019ApJ...881..113C}; therefore, we consider here
$\mcr\in\{\num{\approx4},\allowbreak\num{\approx8},\allowbreak
\num{\approx16},\allowbreak\num{\approx32}\}$, which are typical for \acp{TDE}
in \acp{AGN}. The last three values of $\mcr$ necessitate new runs not reported
in \citet{2019ApJ...881..113C}.

We characterize the \ac{AGN} in terms of its unperturbed disk luminosity
$\eta\amc c^2$, where $\eta=0.1$ is the fiducial accretion efficiency. Every
value of $\mcr$ above the minimum corresponds to two values of $\amc$, the
larger and smaller values appropriate for a disk whose pressure is dominated by
radiation and gas, respectively \citep{2019ApJ...881..113C}. We adopt the
larger value here.

\section{Second impact}
\label{sec:second impact}

\subsection{Trajectory calculations}

\Cref{fig:stream} combines a density slice from the actual simulations with a
schematic depiction of the bigger picture. A stream with $\mcr\gtrsim1$ punches
through the disk at the first impact. Tidal compression redistributes energy
within the stream; as a result, a dilute, wide-angle plume of stream material
bursts out from the lower side of the disk. Our focus is on the fraction that
returns to the disk.

Because stream material leaves the simulation domain with sound speeds much
less than virial, $\mathrelp\lesssim0.1\,[G\Mh/(R^2+z^2)^{1/2}]^{1/2}$
\citep{2019ApJ...881..113C}, we assume it follows ballistic trajectories from
the boundaries of the simulation domain until they strike the disk. For each
time step in each run, we determine which boundary cells have outflowing gas
with specific angular momentum $\mathrelp\ge1.1\,(G\Mh\rp)^{1/2}$ in the
$y$\nobreakdash-direction; these cells have outflowing stream material, as
opposed to vertically expanding disk gas. For each of these cells, we assume
that all the outflowing material over the time step is launched outward on a
Keplerian trajectory. We calculate when, where, and with what velocity this
trajectory hits the midplane. Even gravitationally unbound gas can be
intercepted if its hyperbolic trajectory crosses the disk. Trajectory
calculations are continued until the end of the simulations, at time
$1400\,(G\Mh/\rp^3)^{-1/2}$, which is more than double the simulation duration
in \citet{2019ApJ...881..113C}; for $\Mh=\SI{3e6}{\solarmass}$ and
$\Ms=\si{\solarmass}$, this translates to \SI{\sim10}{\day}, or about a quarter
of the orbital period of the most bound debris when these parameters apply. The
mass fallback rate as a function of time is a convolution of the outflow rate
and the flight-time delay.

\subsection{Dynamics and energetics}
\label{sec:dynamics}

\begin{figure}
\includegraphics{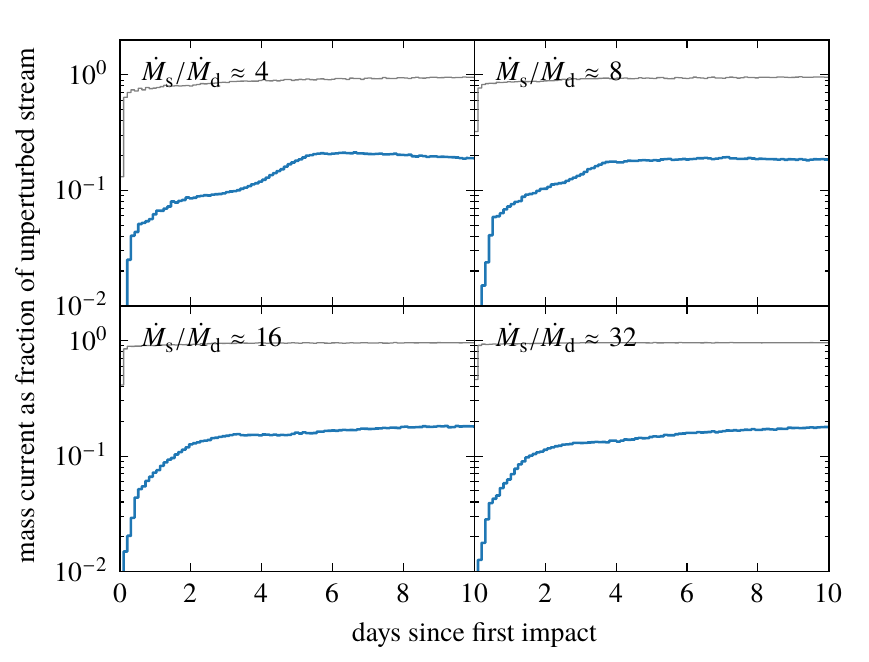}
\caption{Rate at which stream material leaves the simulation domain as a thin
curve, and rate at which it falls back to the midplane as a thick curve, for
$\Mh=\SI{3e6}{\solarmass}$ and $\Ms=\si{\solarmass}$. All rates are normalized
by the unperturbed stream mass current $\smc$.}
\label{fig:mass current}
\end{figure}

\cref{fig:mass current} contrasts the rate at which stream material crashes
back to the disk as a function of time with the rate at which it exits the
first-impact region. Most of the material is unbound, and only
\SI{\sim20}{\percent} is involved in the second impact.

\begin{figure*}
\ifapj\includegraphics{heatmap2}\else
\begin{tikzpicture}[black, nodes={font=\footnotesize}]
\pgfmathsetmacro\xl{-45}
\pgfmathsetmacro\xu{2}
\pgfmathsetmacro\yl{-3}
\pgfmathsetmacro\yu{3}
\pgfmathsetmacro\u{2.8in/(\xu-\xl)}
\pgfmathsetmacro\rc{8}
\pgfmathsetmacro\rd{.2ex/\u}
\node[inner sep=0ex, below right] {\includegraphics{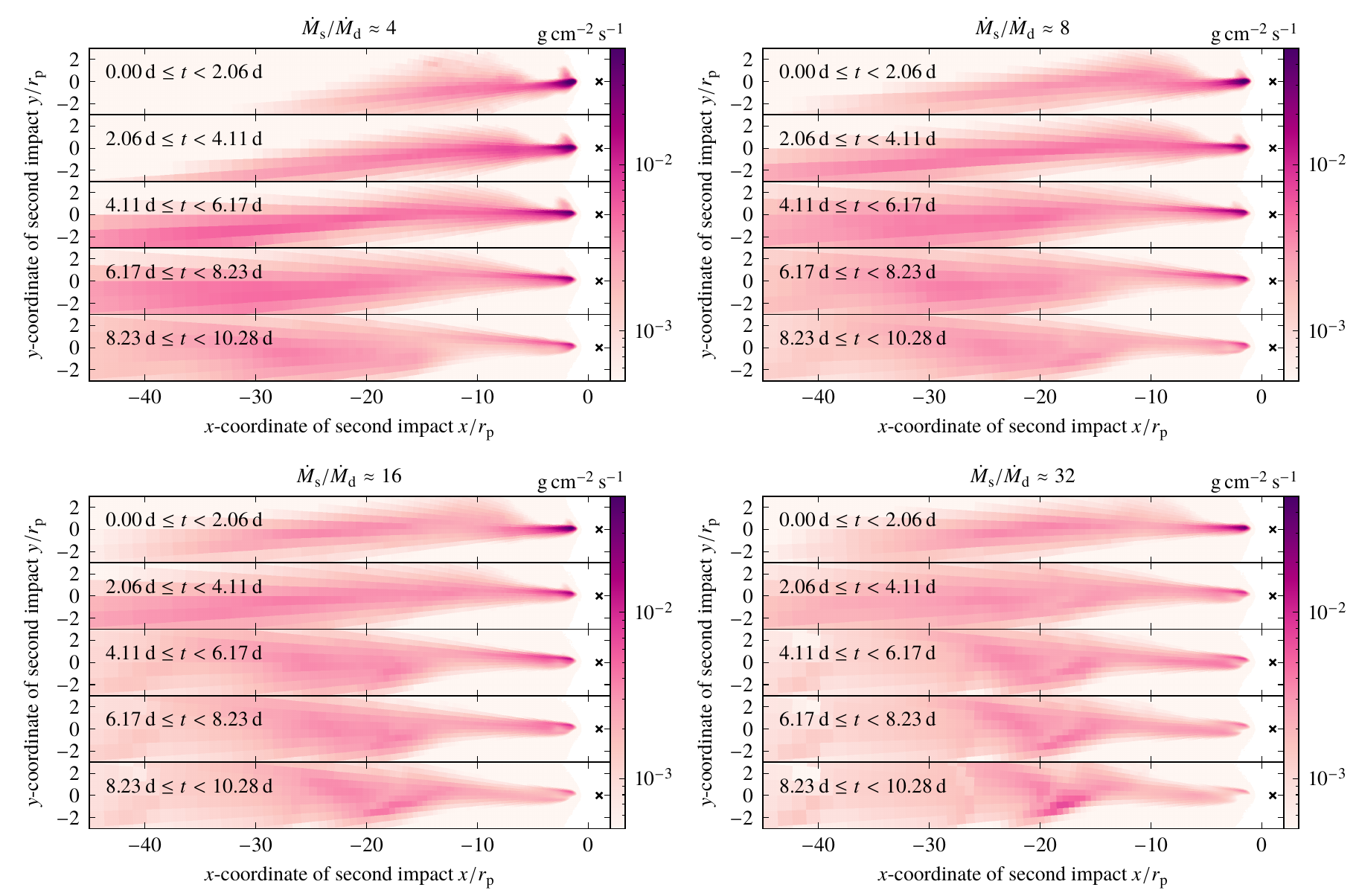}};
\begin{scope}[x=\u, y=\u]
\draw[densely dashed]
  (.48in,-.26in) ++(-\xl,-\yu) +({180-asin(\yu/(\rc-\rd))}:\rc-\rd)
  arc ({180-asin(\yu/(\rc-\rd))}:{180-asin(\yl/(\rc-\rd))}:\rc-\rd);
\draw[densely dashed]
  (.48in,-.26in) ++(-\xl,-\yu) +({180-asin(\yu/(\rc+\rd))}:\rc+\rd)
  arc ({180-asin(\yu/(\rc+\rd))}:{180-asin(\yl/(\rc+\rd))}:\rc+\rd);
\draw[<-]
  (.48in,-.26in) ++(-\xl,-\yu) ++({180-asin(\yu/\rc}:\rc)
  ++(0,.5ex) -- +(0,2ex)
  node [above, align=center] {focus of\\\cref{fig:shocked}};
\end{scope}
\end{tikzpicture}
\fi
\caption{Flux of stream material falling back to the midplane for
$\Mh=\SI{3e6}{\solarmass}$ and $\Ms=\si{\solarmass}$. The flux is the
time integral of mass landing in a cell over one of the five intervals
indicated in each panel, divided by the area of the cell and the length of the
interval. The black hole is at the origin, and the cross marks the first
impact. \Cref{fig:shocked} shows how the second impact along an arbitrary
annulus of the disk, such as the dashed one in the top-left panel, looks like
as viewed from the black hole.}
\label{fig:heatmap}
\end{figure*}

\begin{figure*}
\includegraphics{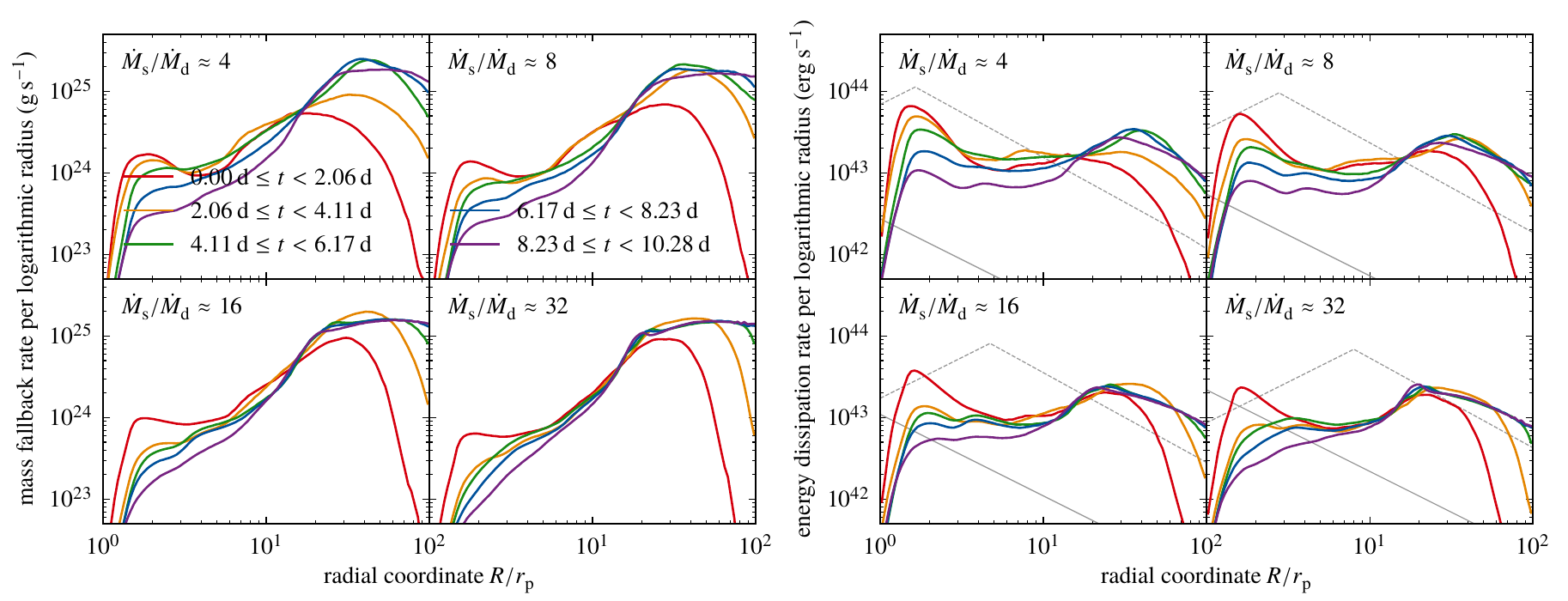}
\caption{\textit{Left panel:} Mass fallback rate per logarithmic radius at the
second impact for $\Mh=\SI{3e6}{\solarmass}$ and $\Ms=\si{\solarmass}$. The
legend applies to the right panel as well. \textit{Right panel:} Energy
dissipation rate per logarithmic radius as given by \cref{eq:energy
dissipation}, for the same $\Mh$ and $\Ms$. The thin solid line is 10 times the
energy dissipation rate per logarithmic radius of the underlying disk. The thin
dashed line is \num{e-3} times the binding energy per logarithmic radius of the
disk divided by the orbital time; the break in this line is at the radius where
disk pressure changes from radiation-dominated on the inside to gas-dominated
on the outside.}
\label{fig:rates}
\end{figure*}

\cref{fig:heatmap} is a map of where the material lands on the disk over the
course of the simulation, and the left panel of \cref{fig:rates} displays the
azimuthally integrated mass fallback rate. The trajectories followed by the
bulk of the material from the first impact to the second have eccentricities
\num{\approx1}, and the planes of the trajectories are almost perpendicular to
the disk. These trajectories dump the material along the projection of the
unperturbed stream onto the disk. A tiny portion of the material is heavily
deflected and put on mildly eccentric, highly inclined trajectories with a
range of orientations; this material ends up at $r\lesssim3\,\rp$. Although the
detailed shape of the splash zone and the rate at which material is delivered
to it vary with time and $\mcr$, overall they are remarkably independent of
these two variables. This means the properties of the second impact depend less
on \ac{AGN} properties ($\amc$) and more on \ac{TDE} properties ($\Mh$, $\Ms$).

The right panel of \cref{fig:rates} shows the azimuthally integrated energy
dissipation rate per logarithmic radius:
\begin{equation}\label{eq:energy dissipation}
R^2\int d\phi\,\rhof\vfz K,
\end{equation}
where $\rhof$ and $\vf$ are the density and velocity of the material falling
back, $K\eqdef\tfrac12\norm{\vf-R\Omega\uvec e_\phi}^2$ is the specific energy
dissipation rate, $\Omega=(G\Mh/R^3)^{1/2}$ is the Keplerian orbital frequency,
and the integral is over where the material hits. At radii greater than a few
$\rp$, material hits the disk with $\norm\vf\ll R\Omega$; as a result, the
dissipated energy is mainly derived from the kinetic energy of the disk, that
is, $K\propto1/R$. At all radii, the energy dissipation rate due to the second
impact is orders of magnitude greater than that of the underlying disk; in
fact, the energy dissipated per orbital time can reach \num{\sim e-3} times the
local binding energy at certain radii. The expedited inflow this entails could
help feed the small-scale disk emptied by the shocks from the first impact
\citep{2019ApJ...881..113C}.

The spatial distribution and energy dissipation of the second impact depend on
the strength of the pericentric tidal compression. This in turn depends on the
kinematic properties of the unperturbed stream, which cannot be accurately
determined without simulating the full disruption process. Therefore, the
results presented here should be understood as capturing the qualitative, not
quantitative, aspects of the second impact.

\section{Compton cooling}
\label{sec:cooling}

The next step is to estimate the photon energies at which the dissipated energy
is radiated from the disk. The task is non-trivial because, at the same time
the stream material shocks and dissipates its energy, its internal energy is
converted to radiation through Compton cooling and carried away with the
escaping radiation. Here we construct a crude model that captures the essence
of these concurrent processes; characterization of the potentially
radiation-dominated shocks and detailed spectral calculations are left to
future work.

\subsection{Overall picture}
\label{sec:overall}

\begin{figure}
\ifapj\includegraphics{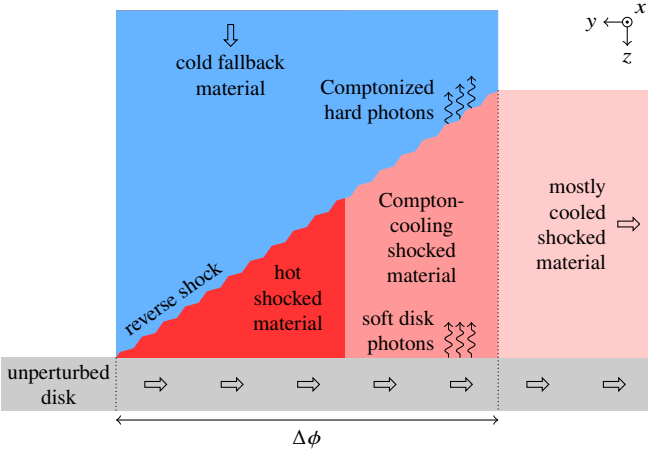}\else
\begin{tikzpicture}[x=2in, y=1.4in, black, nodes={font=\footnotesize}]
\pgfmathsetmacro\a{0.3}
\pgfmathsetmacro\b{1.4}
\pgfmathsetmacro\c{0.2}
\pgfmathsetmacro\d{1.3}
\pgfmathsetmacro\e{0.6}
\fill[black!20]
  (-\a,0) rectangle (\b,-\c);
\fill[red!80]
  (0,0) rectangle (\e,\d);
\fill[red!40]
  (\e,0) rectangle (1,\d);
\fill[blue!50!cyan!60]
  (0,0) decorate [decoration={zigzag, amplitude=.2ex}] {-- (1,1)} --
  (1,\d) -- (0,\d) -- cycle;
\fill[red!20]
  (1,0) rectangle (\b,1);
\draw[densely dotted]
  (0,0) -- (0,-\c) (1,1) -- (1,-\c);
\node at (-\a/2,-\c/2)
  [align=center] {unperturbed\\disk};
\path (0,0) -- (1,1)
  node [pos=0.03, above right, sloped] {reverse shock};
\node at (\e*3/4,\e*3/8)
  [align=center] {hot\\shocked\\material};
\node at ({(\e+1)/2},{(\e+1)/4+0.05})
  [align=center] {Compton-\\cooling\\shocked\\material};
\begin{scope}[flow/.style={draw, single arrow, text width=1.5ex,
  text height=.4ex, inner sep=0ex, single arrow head extend=.3ex}]
\foreach \i in {0.1,0.3,...,\b} \node at (\i,-\c/2) [flow] {};
\node at ([yshift=-2ex] 0.3,\d) [flow, rotate=-90] {};
\node at ([yshift=-3ex] 0.3,\d)
  [below, align=center] {cold fallback\\material};
\node at ([xshift=-2ex] \b,0.5) [flow] {};
\node at ([xshift=-3ex] \b,0.5)
  [left, align=right] {mostly\\cooled\\shocked\\material};
\end{scope}
\begin{scope}[shift={({(\e+3)/4},0)}]
\foreach \x in {-0.03,0,0.03}
  \pic[rotate=90] at (\x,0) {photon};
\node at (-0.03,1.5ex) [left=.4ex, align=right, yshift=.6ex]
  {soft disk\\photons};
\end{scope}
\begin{scope}
\clip
  (0,0) decorate [decoration={zigzag, amplitude=.2ex}] {-- (1,1)} --
  (1,\d) -- (0,\d) -- cycle;
\begin{scope}[shift={({(\e+3)/4},{(\e+3)/4})}, yshift=-.2ex]
\foreach \x in {-0.03,0,0.03}
  \pic[rotate=90] at (\x,\x) {photon};
\node at (-0.03,1.5ex) [left=.4ex, align=right] {Comptonized\\hard photons};
\end{scope}
\end{scope}
\draw[<->, yshift=-.75ex] (0,-\c) -- (1,-\c)
  node [midway, below] {$\Delta\phi$};
\pic[shift={(-2ex,-1ex)}, rotate=180] at (current bounding box.north east)
  {triad out=yzx};
\end{tikzpicture}
\fi
\caption{Schematic diagram of the second impact. This figure shows how the
skinny dashed box in \cref{fig:stream} or the dashed annulus in
\cref{fig:heatmap} looks from the perspective of the black hole; colors
represent in an approximate manner the different physical states of the gas.
The unperturbed disk (gray) moves in from the left. Cold stream material (blue)
falls inelastically onto the disk and is constantly transported to the right by
disk rotation. The standing, oblique reverse shock thus formed heats the
material (dark red). At the same time, soft seed photons from the disk undergo
inverse Compton scattering and harden as they propagate upward through the
shocked material. Compton cooling is efficient only in parts of the material
(light red) that are somewhat Compton-thick in the vertical direction. This
mostly cooled material (pale red) likely sinks into the disk over the course of
an orbit, so that the disk rotates back into the stream in its unperturbed
state.}
\label{fig:shocked}
\end{figure}

\Cref{fig:shocked} shows an idealized picture of the second impact in the
inertial frame. Falling stream material is slowed down, compressed, and heated
by a strong, standing reverse shock. The layer of shocked material, sitting on
the disk, is ferried away by disk rotation. This material cools quite swiftly;
with the loss of pressure support, presumably it sinks into the disk within one
orbit. As is evident from the narrowness of the splash zones displayed in
\cref{fig:heatmap}, the time for the disk to rotate through the stream is much
shorter than an orbital period. The structure in \cref{fig:shocked} is thus
time-steady: the unperturbed disk enters from the left side, and exits to the
right topped with cooled shocked material.

Over longer timescales, this time-steady picture changes gradually as the rate
at which stream material descends on the disk varies. The mass current of the
second impact slowly builds up in the early stages of the \ac{TDE} because the
stream takes time to fly from the first impact to the second. In addition, on
timescales comparable to the mass-return time of the \ac{TDE}, the rate at
which stellar material returns to pericenter declines, and so does the mass
current going to the second impact.

Our simulations, which last for \SI{\sim10}{\day} when
$\Mh=\SI{3e6}{\solarmass}$ and $\Ms=\si{\solarmass}$, are long enough to cover
only the buildup period. To study how the second impact changes over this
period, we divide the simulation duration into five equal intervals. The
fallback material is characterized by its speed, mass flux, and energy
dissipation rate; for simplicity, we assume these three quantities are
time-independent within each interval and $\phi$\nobreakdash-independent across
the stream, so we can consider their averages $\mean\vfz$, $\mean{\rhof\vfz}$,
and $\mean{\rhof\vfz K}$.

\subsection{Cooling model}
\label{sec:model}

To calculate the cooling emission, we follow a point on the disk at radius $R$
from the moment it enters the stream, which we take to be $t=0$. A column of
shocked material accumulates above the point as it moves through the stream;
meanwhile, the column cools by radiating away the dissipated energy. The point
leaves the stream at $\tdep=\Delta\phi/\Omega$, where $\Delta\phi$ is the
characteristic azimuthal width of the second impact at radius $R$; we
nevertheless continue tracking the cooling over the entire orbit to ensure the
column has largely cooled off by the time it returns to the stream. The
observed cooling emission is the sum from columns all over the disk, each in a
different cooling stage.

The dominant cooling mechanism is inverse Compton scattering of seed photons
from the disk. In this sense, the shocked layer behaves similarly to \iac{AGN}
corona \citep[e.g.,][]{1991ApJ...380L..51H}, except that the shocked layer can
be quite Compton-thick. Synchrotron emission should not contribute greatly
because it is strongly self-absorbed, and we shall see later that free--free
cooling is also unlikely to be important.

We assume the shocked layer is vertically homogeneous, and ions and electrons
are thermal at the same temperature. Thermal electrons at the temperatures we
find are mildly relativistic; therefore, we ignore electron--positron pair
production and the Klein--Nishina reduction in cross section, both affecting
only a small fraction of photons.

The spectrum of the Comptonized photons is taken to be a power law with a
cutoff. Seed photons from the disk carry a radiative flux
\begin{equation}
\Fd=\frac{3G\Mh\amc}{8\pi R^3},
\end{equation}
and their spectrum is black-body at temperature $\Td=(\Fd/\sigmaSB)^{1/4}$,
where $\sigmaSB$ is the Stefan--Boltzmann constant. Because the seed spectrum
is much narrower than the Comptonized spectrum, we approximate the former as a
delta function. We use $\Fd$ of the unperturbed disk here, but spiral shocks
extending outward from the first impact \citep{2019ApJ...881..113C} could
modify the disk and thus $\Fd$.

At each moment, the model is characterized by $\Eg$ and $\Er$, which are,
respectively, the vertically integrated gas and radiation energy densities; by
$\Einfty$, the energy per area that has escaped from the top of the column; and
by $\Hsh$, the height of the column of shocked material. They obey the
equations
\begin{align}
\label{eq:model gas}
\od\Eg t &= \mean{\rhof\vfz K}\step{\tdep-t}-(A-4\ThetaC)\frac\Er\tsca, \\
\label{eq:model radiation}
\od\Er t &= \Fd+(A-4\ThetaC)\frac\Er\tsca-\frac\Er\tesc, \\
\od\Einfty t &= \frac\Er\tesc, \\
\label{eq:model height}
\od\Hsh t &= \begin{cases}
  \vsh & \mtext{if $t<\tdep$,} \\
  (\Heq/\Hsh-1)\cs & \mtext{if $t\ge\tdep$,}
\end{cases}
\end{align}
where $\vartheta$ is the step function, and the other symbols are defined in
the rest of the section. The initial condition is
$(\Eg,\Er,\Einfty,\Hsh)=(0,0,0,0)$; the results are insensitive to small
perturbations to the initial condition.

\Cref{eq:model gas} characterizes the accrual of dissipated energy in the
column and its conversion to radiation through Compton scattering.
\Cref{eq:model radiation} follows the increase in radiation energy as a result
of disk injection and Compton scattering, and its decrease due to radiative
losses. \Cref{eq:model height} describes how the column grows during deposition
while mass and heat are added to it, and collapses after deposition as it
cools. We ignore the internal energy increase during this collapse due to
adiabatic compression and release of gravitational energy, both being secondary
effects.

In \cref{eq:model gas,eq:model radiation}, $A$ is the fractional photon energy
gain due to inverse Compton scattering, and $\ThetaC$ is the Compton
temperature in units of $\me c^2/\kB$, with $\me$ the electron mass and $\kB$
the Boltzmann constant. The Compton temperature is the temperature of an
electron gas in thermal balance with photons through Compton scattering, and it
is equal to one-fourth the intensity-weighted mean photon energy.

Three timescales appear in the equations:
\begin{align}
\tdep &= \Delta\phi/\Omega, \\
\tsca &= 1/(\nelc\sigmaT c), \\
\tesc &= (\Hsh/c)\max(1,\tauT),
\end{align}
where $\nelc$ is the electron number density, $\sigmaT$ is the Thomson cross
section, and $\tauT$ is the Thomson thickness. These timescales are,
respectively, the duration of deposition, the mean timescale for a photon to
scatter off an electron, and the timescale for radiation to escape. We also
define
\begin{align}
\vsh &= [(\gamma-1)/(\gamma+1)]\mean\vfz, \\
\Heq^2 &= 2\cs^2R^3/(G\Mh), \\
\cs^2 &= (\tfrac23\Eg+\tfrac13\Er)/\Sigmash, \\
\nelc &= \Sigmash/(2\bar m\Hsh), \\
\tauT &= \sigmaT\Sigmash/\mH, \\
\Sigmash &= \mean{\rhof\vfz}\min(\tdep,t).
\end{align}
Here $\vsh$ is the upward speed of the reverse shock in the corotating frame,
$\Heq$ is the scale height in hydrostatic equilibrium, $\cs$ is the isothermal
sound speed, $\Sigmash$ is the column density, $\gamma=\tfrac53$ is the
adiabatic index, $\bar m=\tfrac8{13}\mH$ is the mean particle mass, and $\mH$
is the hydrogen mass.

Although the electrons are mildly relativistic at the beginning, the radiative
output is dominated by later stages when the gas is cooler; we therefore have
\begin{equation}
A\approx4\Theta,
\end{equation}
where
\begin{equation}
\Theta\eqdef\kB\Tg/(\me c^2),
\end{equation}
and
\begin{equation}
\Tg=(\gamma-1)\bar m\Eg/(\kB\Sigmash)
\end{equation}
is the gas temperature.

We compute the Compton temperature in two different ways, distinguished by
whether the column is Compton-thin or thick. The dividing line $\tauT=1/e$ is
chosen to be as large as possible while being small enough that the probability
of $n$ scatters is $\mathrelp\approx\tauT^n$. If $\tauT<1/e$, most disk photons
escape without being scattered, so the spectrum of the radiation emerging from
the column consists of a delta-function disk component in addition to a
Comptonized component:
\begin{multline}
F(\epsilon)\,d\epsilon=\biggl[(1-\tauT)\Fd\dfunc{\epsilon-\kB\Td} \\
+\frac\FC{\kB\Tg}
  \biggl(\frac\epsilon{\kB\Tg}\biggr)^p
  \exp\biggl(-\frac\epsilon{\kB\Tg}\biggr)\biggr]\,d\epsilon,
\end{multline}
where $\epsilon$ is the photon energy and $\FC$ is the normalization of the
Comptonized component. In this regime, repeated Compton scattering leads to a
spectral index $p=\ln\tauT/\ln(1+A)$ \citep{1986rpa..book.....R,
1999agnc.book.....K}, and energy conservation and the definition of $\ThetaC$
can be written as
\begin{align}
\frac\Er\tesc
  &= \int_{\kB\Td^-}^\infty d\epsilon\,F(\epsilon) \nonumber\\
&= (1-\tauT)\Fd+\Gamma(p+1,\Td/\Tg)\FC, \\
4\ThetaC\me c^2
  &= \frac
  {\int_{\kB\Td^-}^\infty d\epsilon\,\epsilon F(\epsilon)}
  {\int_{\kB\Td^-}^\infty d\epsilon\,F(\epsilon)} \nonumber\\
&= \frac
  {(1-\tauT)(\Td/\Tg)\Fd+\Gamma(p+2,\Td/\Tg)\FC}
  {(1-\tauT)\Fd+\Gamma(p+1,\Td/\Tg)\FC}\kB\Tg,
\end{align}
with $\Gamma$ the incomplete gamma function. Solving these equations yields
$\FC$ and $\ThetaC$; in case $\FC<0$, we set $\ThetaC=0$.

If $\tauT\ge1/e$, we assume every photon is scattered at least once on its way
out, so we retain only the Comptonized component:
\begin{equation}
F(\epsilon)\,d\epsilon=\frac\FC{\kB\Tg}
  \biggl(\frac\epsilon{\kB\Tg}\biggr)^p
  \exp\biggl(-\frac\epsilon{\kB\Tg}\biggr)\,d\epsilon.
\end{equation}
The spectral parameters $\FC$, $p$, and $\ThetaC$ are determined in this regime
from photon conservation, energy conservation, and the definition of $\ThetaC$:
\begin{alignat}{2}
\frac\Fd{\kB\Td}
  &= \int_{\kB\Td}^\infty d\epsilon\,\epsilon^{-1}F(\epsilon)
  &&= \Gamma(p,\Td/\Tg)\frac\FC{\kB\Tg}, \\
\frac\Er\tesc
  &= \int_{\kB\Td}^\infty d\epsilon\,F(\epsilon)
  &&= \Gamma(p+1,\Td/\Tg)\FC, \\
4\ThetaC\me c^2
  &= \frac
  {\int_{\kB\Td}^\infty d\epsilon\,\epsilon F(\epsilon)}
  {\int_{\kB\Td}^\infty d\epsilon\,F(\epsilon)}
  &&= \frac{\Gamma(p+2,\Td/\Tg)}{\Gamma(p+1,\Td/\Tg)}\kB\Tg.
\end{alignat}

\subsection{Solution}
\label{sec:solution}

\begin{figure}[!t]
\includegraphics{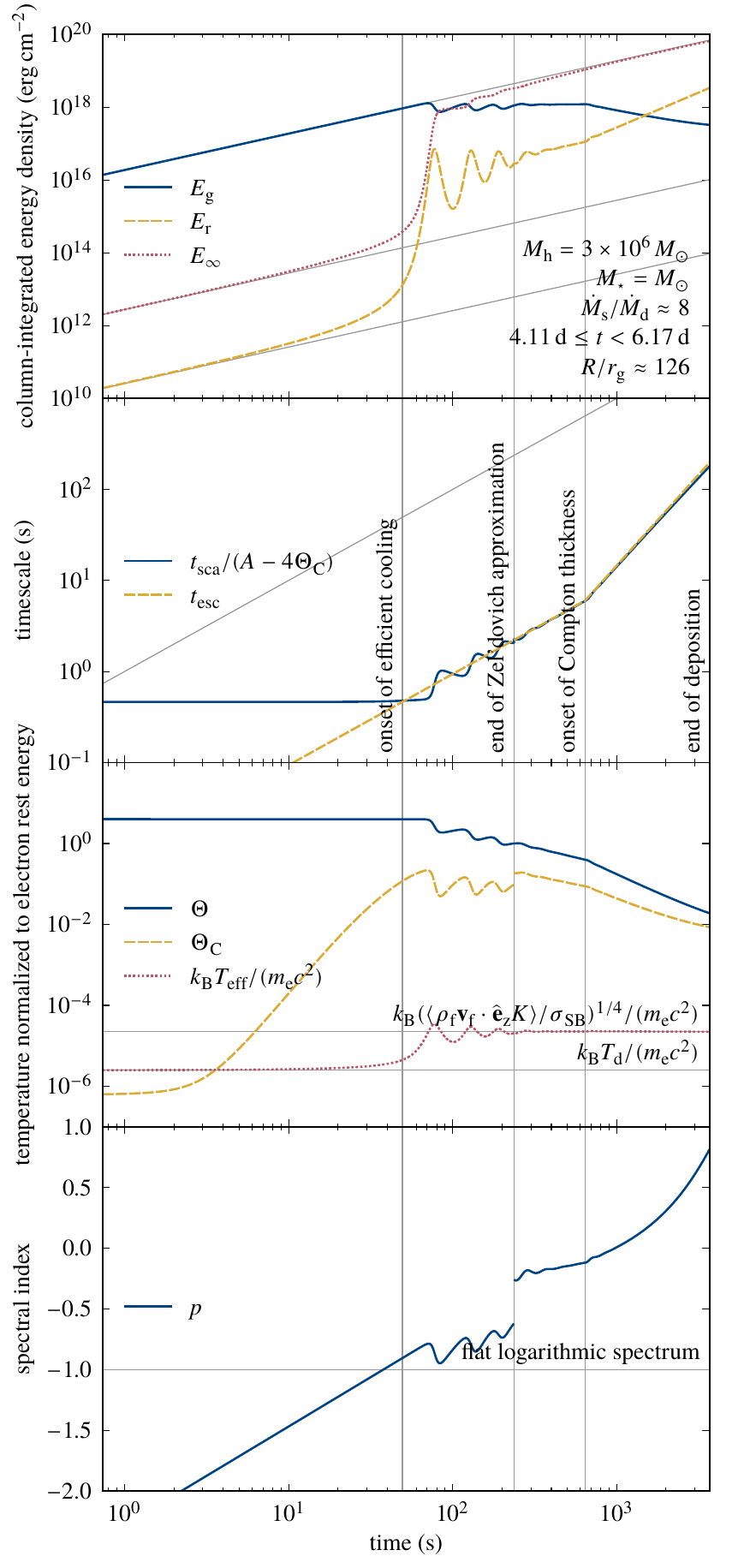}
\caption{One solution of the cooling model; see \cref{sec:model} for
definitions. The vertical lines from left to right mark the times when
$(A-4\ThetaC)\tesc/\tsca=1$, $\tauT=1/e$, and $\tauT=1$, respectively; the
first two times may be interchanged in other solutions. The inclined lines in
the first panel are given by \cref{eq:solution gas,eq:solution
radiation,eq:solution escaped}. The inclined line in the second panel follows a
one-to-one ratio.}
\label{fig:model}
\end{figure}

\Cref{fig:model} shows one solution of the model; other solutions are
qualitatively similar, differing only in the durations of the various stages of
evolution. The first stage is so optically thin that
$(A-4\ThetaC)\tesc/\tsca=(A-4\ThetaC)\tauT\mH/(2\bar m)\ll1$; thus, the net
amplification of the disk emission is small. The third term on the right-hand
side of \cref{eq:model radiation} largely offsets the first term, so $\Er$
grows slowly and $\Tg$ remains high. Indeed,
\begin{align}
\label{eq:solution gas}
\Eg &\approx \mean{\rhof\vfz K}t, \\
\label{eq:solution radiation}
\Er &\approx \Fd t/(1+c/\vsh), \\
\label{eq:solution escaped}
\Einfty &\approx \Fd t.
\end{align}

The second stage commences when
$(A-4\ThetaC)\tesc/\tsca=(A-4\ThetaC)\tauT\mH/(2\bar m)\sim1$. Although the
column is still Compton-thin, the scattered photons gain so much energy that
the luminosity of a cohort of injected photons is at least doubled. This drives
up $\Er$, which in turn accelerates Compton cooling, resulting in an
exponential surge in $\Er$. The oscillations in this stage are due to the
mutual feedback between $\Eg$ and $\Er$ in our model, in which a column evolves
independently of its neighbors; if adjacent columns interacted by photon
diffusion, these oscillations would likely be damped.

The growth of $\Er$ comes to an end when the system finds a new equilibrium. In
this equilibrium, dissipated energy is efficiently converted to radiation
energy, which in turn is lost rapidly to outward streaming. This means the two
terms of \cref{eq:model gas} are comparable, while the second term of
\cref{eq:model radiation} is counteracted by the third. The effective
temperature of the emergent radiation is then
\begin{equation}
\Teff=\biggl(\frac\Er{\tesc\sigmaSB}\biggr)^{1/4}
  \sim\biggl(\frac{\mean{\rhof\vfz K}}\sigmaSB\biggr)^{1/4}.
\end{equation}

The third stage goes from the moment when $\tauT\sim1$, should it happen, to
the end of deposition. Compton cooling is faster in the Compton-thick column,
but the gas is kept warm by continual energy dissipation. The balance of terms
is the same as in the second stage; hence,
$(A-4\ThetaC)\tesc/\tsca=(A-4\ThetaC)\tauT^2\mH/(2\bar m)\sim1$. The close
match between dissipative gains and radiative losses means that, once
deposition ceases, the gas promptly cools off and little further emission
occurs.

Because all quantities as functions of time are approximately power-law, later
stages of cooling tend to outweigh earlier stages in terms of contribution to
the overall radiative output.

On the whole, both $p$ and $\ThetaC$ increase over time in all three stages as
long as material continues to be deposited, but $p$ never rises to above 3, its
value in the Rayleigh--Jeans limit. By contrast, $\Theta$ is almost constant in
the first stage, but falls gradually in the next two stages, bounded below by
$\ThetaC$. This observation can be further generalized: a Compton-thicker layer
tends to have a larger $p$ and a smaller $\Theta$. Consequently, the spectrum
emerging from the column transforms from a soft power law with a high-energy
cutoff to a hard power law with a low-energy cutoff as cooling progresses.

Free--free cooling is more efficient for small $\Mh$, large $\Ms$, and large
$R/\rp$. It can be orders of magnitude faster than Compton cooling at
converting $\Eg$ to $\Er$ in the first phase, but for most of the cases we
considered, Compton cooling quickly catches up in the second and third phases,
and even monopolizes the cooling budget toward the end of deposition.
Therefore, the effect of early-time free--free cooling is likely limited to
creating extra seed photons for Compton cooling. Free--free cooling can also
dominate after the end of deposition; however, in most cases only a small
fraction of the total dissipated energy remains in the column by the time
stream material stops arriving, so the contribution of late-time free--free
cooling is minor.

Another source of seed photons is the disk interior to the first impact, which
could be emitting at Eddington levels \citep{2019ApJ...881..113C,
2020ApJ...903...17C}. Although this could, in principle, be a competitive
source of seed photons, the luminosity and angular distribution of this light
remain so uncertain that we do not include it here.

\subsection{Spectra and luminosities}
\label{sec:spectra}

\begin{figure*}
\includegraphics{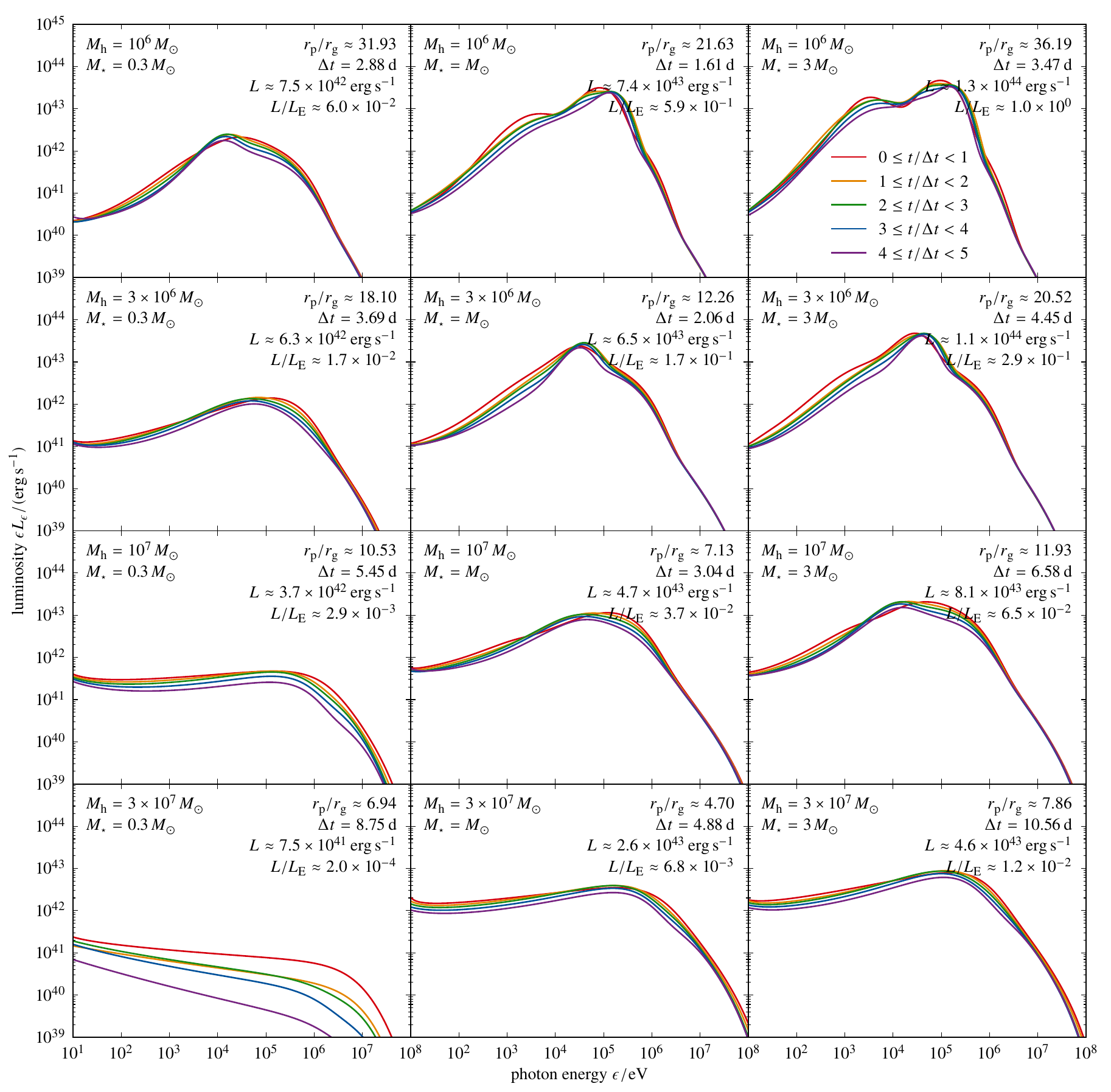}
\caption{Disk-integrated second-impact spectra for $\mcr\approx8$. All panels
share the same legend, but the value of $\Delta t=280\,(G\Mh/\rp^3)^{-1/2}$
varies from panel to panel. Here $\LE=4\pi G\Mh c\mH/\sigmaT$ is the Eddington
luminosity.}
\label{fig:spectrum}
\end{figure*}

The analysis so far concerns a single column. \Cref{fig:spectrum} shows the
overall spectra resulting from the second impact for various values of $\Mh$
and $\Ms$. These are obtained by adding together the contribution of every
column on the disk. In all cases, it can be roughly represented by a broken
power law whose break coincides with the peak in $\epsilon L_\epsilon$.
Depending on $\Mh$ and $\Ms$, the photon energy of the peak can be anywhere
between \SI{\sim10}{\kilo\electronvolt} and \SI{1}{\mega\electronvolt}, as
expected from Compton cooling of mildly relativistic electrons. Generally
speaking, with increasing $\Mh$ and decreasing $\Ms$, the peak broadens and
shifts toward higher energies. Additional cooling at $t\gtrsim\tdep$ modifies
the spectrum, the most prominent effect being the creation of a secondary
$\mathrelp\sim\si{\kilo\electronvolt}$ peak for small $\Mh$ and large $\Ms$.

\begin{figure}
\includegraphics{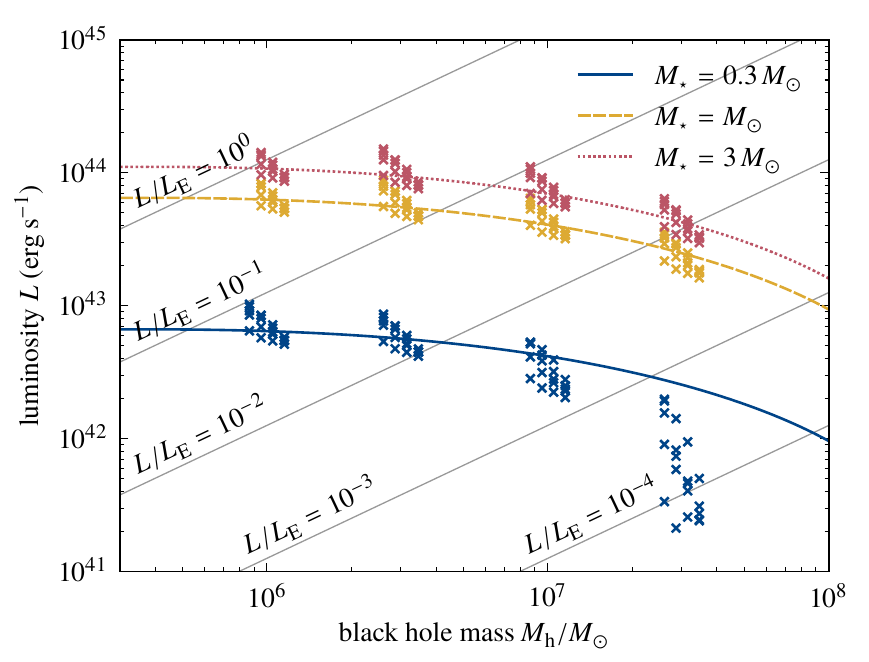}
\caption{Luminosity of the second impact. The crosses are the luminosities
predicted by our cooling model; there is one cross for every $\mcr$ and for
every one-fifth of the simulation duration (\cref{sec:overall}). For clarity,
the crosses are displaced horizontally by a small amount according to their
$\mcr$, with $\mcr$ increasing from left to right. The luminosity depends
weakly on $\mcr$ and time in most cases. The curves are given by
\cref{eq:luminosity}. That equation estimates the second-impact luminosity to
be equal to its characteristic energy dissipation rate, and the overall
normalization of the estimate is chosen to fit the model luminosities. The
goodness of the fit suggests that the material shocked at the second impact
cools efficiently.}
\label{fig:scaling}
\end{figure}

The spectra in \cref{fig:spectrum} have bolometric luminosities ranging from
\SIrange{\sim e42}{e44}{\erg\per\second}, on the level of weak \acp{AGN}.
Because the shocked material cools efficiently, these luminosities are close to
the total energy dissipation rate, which is $\mathrelp\sim0.2\,\smc$ times the
specific kinetic energy of the disk at the second impact (\cref{sec:dynamics}).
Second-impact radii are multiples of $\rp$, so the luminosity scales as
\begin{align}
L\propto G\Mh\smc/\rp
  &\propto \Mh^{1/6}\Ms^{7/3}\rs^{-5/2}\Psi^{-1}\Xi^{3/2} \nonumber\\
&\propto \Mh^{1/6}\Ms^{0.13}\Psi^{-1}\Xi^{3/2}.
\end{align}
As seen in \cref{fig:scaling}, this equation best describes the luminosities
from the model when the constant associated with the first proportionality is
\num{\sim4e-3}:
\begin{equation}\label{eq:luminosity}
L\sim\SI{1.8e43}{\erg\per\second}\times
  \Psi^{-1}\Xi^{3/2}
  \biggl(\frac\Mh{\SI{3e6}{\solarmass}}\biggr)^{1/6}
  \biggl(\frac\Ms{\si{\solarmass}}\biggr)^{0.13}.
\end{equation}
The proportionality constant is so small because merely \SI{\sim20}{\percent}
of the bound debris participates in the second impact, and because, according
to \cref{fig:heatmap}, the second impact happens at several tens of $\rp$.

In view of the weak \textquote{nominal} dependence of $L$ on $\Mh$ and $\Ms$,
the correction factors $\Psi$ and $\Xi$ actually control how $L$ scales with
$\Mh$ and $\Ms$. The combination $\Psi^{-1}\Xi^{3/2}$ decreases with $\Mh$ and
increases with $\Ms$, counteracting the nominal $\Mh$\nobreakdash-dependence
but augmenting the nominal $\Ms$\nobreakdash-dependence. As is apparent from
\cref{fig:spectrum,fig:scaling}, $L$ is most sensitive to $\Mh$ for
$\Mh\gtrsim\SI{3e6}{\solarmass}$, and to $\Ms$ for
$\SI{0.1}{\solarmass}\lesssim\Ms\lesssim\si{\solarmass}$.

\section{Discussion}
\label{sec:discussion}

\subsection{Identifying second-impact signatures in \ac{TDE} searches}

\Acp{TDE} in \acp{AGN} have observational signatures that are emphatically
unlike \acp{TDE} in vacuum. They do not have optical/\ac{UV} light curves that
follow the mass-return rate, as is commonly expected for their vacuum
counterparts \citep[e.g.,][]{1988Natur.333..523R}. Instead, the destruction of
the inner disk by the first impact sustains an Eddington-level luminosity
plateau over tens of days \citep{2019ApJ...881..113C, 2020ApJ...903...17C}. On
top of that, the second impact begets a longer-lasting signal whose luminosity
does follow the mass-return rate thanks to efficient cooling, but its typical
photon energy is not in the optical/\ac{UV}: most of the spectra in
\cref{fig:spectrum} peak between \SI{\sim10}{\kilo\electronvolt} and
\SI{1}{\mega\electronvolt}. Such spectra bear no resemblance to the
tens-of-\si{\electronvolt} thermal spectrum originally expected from the
accretion disk formed by \acp{TDE} in vacuum \citep{1990ApJ...351...38C,
1999ApJ...514..180U} and is sometimes observed in soft \xray{} \acp{TDE}
\citetext{see \citealp{2020SSRv..216...85S} for a review}.

The hard second-impact spectrum could mean that \ac{TDE} searches focused on
softer spectra would miss it altogether. The hunt is made difficult by the fact
that, unlike jetted \acp{TDE} with isotropic-equivalent \xray{} luminosities of
\SIrange{\sim e46}{e48}{\erg\per\second} \citep{2011Sci...333..203B,
2011Natur.476..421B, 2011Sci...333..199L, 2012ApJ...753...77C}, the second
impact emits at rates four orders of magnitude lower.

\subsection{Second-impact signatures versus \ac{AGN} variability}

Another challenge in the identification of second-impact signatures is
distinguishing them from common AGN variability. Two aspects may provide the
means to do so: spectral hardness and pattern of time variation.

The second impact has much harder spectra than ordinary \acp{AGN}. The coronal
component of unobscured \ac{AGN} spectra typically has $-1\lesssim\ods{\log
L_\epsilon}{\log\epsilon}\lesssim-0.7$ from
\SIrange{\sim1}{100}{\kilo\electronvolt} \citep[e.g.,][]{2016MNRAS.459.1602L,
2017ApJS..233...17R}. Judging by \cref{fig:spectrum}, the \xrays{} from the
second impact should have $-1\lesssim\ods{\log
L_\epsilon}{\log\epsilon}\lesssim0$. The contrast between unobscured \ac{AGN}
spectra and the significantly harder second-impact spectra is enhanced to the
degree that the corona is disrupted by the first impact, an outcome suggested
in our earlier work \citep{2019ApJ...881..113C, 2020ApJ...898L...1R}.

In addition to unobscured \acp{AGN}, there is also a roughly comparable number
of obscured \acp{AGN} \citep[e.g.,][]{1992ApJ...393...90H, 2008AJ....136.2373R,
2011MNRAS.414.3084B, 2013ApJ...773...15W, 2015ApJS..219....1O}. Our sightlines
to these \acp{AGN} are obscured by neutral gas with hydrogen column densities
$\NH$ of \SIrange{\sim e22}{e24}{\per\centi\meter\squared} or more. A column
with $\NH\sim\SI{e22}{\per\centi\meter\squared}$ absorbs essentially all
photons below \SI{\sim2}{\kilo\electronvolt}, and one with
$\NH\sim\SI{e24}{\per\centi\meter\squared}$ absorbs all photons under
\SIrange{\sim7}{10}{\kilo\electronvolt}. A column with still higher $\NH$
scatters all photons with energies up to \SI{\gtrsim100}{\kilo\electronvolt}. A
significant fraction of all obscured \acp{AGN}, from
\SIrange{\sim20}{50}{\percent}, may belong to this last category
\citep{2015ApJ...815L..13R, 2019A&A...621A..28G, 2020ApJ...901..161K}. Because
this obscuration is mostly located parsecs away from the black hole, if it
blocks \xrays{} from \iac{AGN}, it will equally block those due to \iac{TDE}.
The only photons we can see are relatively high-energy, which makes the
contrast between the second impact and the unperturbed \ac{AGN} especially
high.

The smooth brightening and fading of the second impact could provide another
method of distinction. The \xray{} light curves of normal \acp{AGN} generically
exhibit \textquote{red noise}, that is, their Fourier power spectra are power
laws declining from timescales of months or years to timescales of hours
\citep[e.g.,][]{1987Natur.325..694L, 1987Natur.325..696M, 1993ApJ...414L..85L},
and individual modes of the power spectra have little or no phase coherence
\citep{1993ApJ...402..432K}. By contrast, the second impact should brighten and
fade smoothly over a period of months, in response to the variation of the
mass-return rate. The existence of a dominant timescale, and the phase
coherence implied by the smoothness of the light curve, may help make a
second-impact flare distinct.

\subsection{Caveats about second-impact signatures}

Our treatment of light production at the second impact depends on a number of
approximations and simplifications. A few warrant identification as possible
starting points for future improvements.

The simulations underlying our cooling model injected the stream on a parabolic
orbit \citep{2019ApJ...881..113C} when, in fact, the debris is weakly bound.
Therefore, the stream's kinetic energy at pericenter is overestimated by an
amount comparable to its typical binding energy. After correcting for this
offset, the second impact would happen at smaller radii. This would increase
the energy dissipated per unit stream mass somewhat, raise the luminosity, and
change the spectrum. A greater fraction of the stream could also end up in the
second impact, making the cooling emission from it softer and more thermal
(\cref{sec:solution}).

As in \citet{2019ApJ...881..113C, 2020ApJ...903...17C}, we consider here a
specific stream configuration: the stream hits the disk perpendicularly while
passing through its pericenter at radius $\Rt$. Our results might change
depending on stream orientation and pericenter, but preliminary estimates
suggest that orientation does not qualitatively affect our conclusions.

Our model ignores the production of electron--positron pairs for simplicity
(\cref{sec:model}), but many pairs should be produced in the freshly shocked
gas with temperatures $\kB\Tg\gtrsim\me c^2$. In pair equilibrium, some of the
thermal energy is held in the rest mass of pairs, while pair annihilation adds
to the number density of photons. The increase in scattering opacity due to
pairs combined with the presence of additional photons tends to promote lower
gas and radiation temperatures, and more nearly thermal spectra.

Lastly, our simulations cover merely a fraction of the mass-return time, during
which the stream mass current can be taken to be approximately constant.
However, as the mass-return rate rises, reaches a peak, and then falls over the
course of the \ac{TDE}, we expect the bolometric luminosity of the second
impact to roughly track that rate, while the spectrum should evolve from harder
at the initially low mass-return rate, to softer and more thermal near the
peak, and then back to harder as the mass-return rate decays.

\section{Conclusions}
\label{sec:conclusions}

As argued in \citet{2019ApJ...881..113C}, some fraction of all \acp{TDE} should
take place in \acp{AGN}. These \acp{TDE} have drastically different physics and
phenomenology from those in inactive galaxies due to the pre-existing \ac{AGN}
disk. The disk does not affect the disruption, but it can block the returning
debris stream. The dynamics and observational properties of these \acp{TDE} are
thus defined by the multiple interactions between the stream and the disk. The
mechanisms commonly considered in connection with \acp{TDE} in vacuum cannot be
applied to \acp{TDE} in \acp{AGN}, and they certainly cannot explain any
optical/\ac{UV} emission from the latter.

For \acp{TDE} in \acp{AGN}, the first impact of the stream with the disk takes
place near stream pericenter. The shocks generated cause the disk interior to
the impact point to fall rapidly into the black hole, and they power a
short-lived Eddington-limited luminosity plateau \citep{2019ApJ...881..113C},
possibly thermal in spectrum. The plateau lasts tens of days
\citep{2020ApJ...903...17C} if the tidal radius is estimated using an
order-of-magnitude estimate \citep[e.g.,][]{1988Natur.333..523R}. Accounting
for the effects of general relativity and stellar structure on the disruption
changes the duration by a factor of a few \citep{2020ApJ...904...98R}; for
example, the plateau duration for a Sun-like star disrupted by a
\SI{3e6}{\solarmass} black hole accreting at 0.01 times Eddington is shortened
from \SI{\sim10}{\day} to \SI{\sim3}{\day}.

In most cases, a large fraction of the stream drives through the disk at the
first impact, only to encounter the disk again at larger distances. The stream
is stopped at this second impact and heated to temperatures
\SI{\sim100}{\kilo\electronvolt}. Compton cooling produces hard \xrays{} to
soft \gammarays{} with a strikingly hard spectrum: most of the energy is
carried by photons from \SI{\sim10}{\kilo\electronvolt} to
\SI{1}{\mega\electronvolt}. The luminosity tracks the mass-return rate,
reaching \SIrange{\sim e42}{e44}{\erg\per\second} at peak; this means the
timescale of the light curve is roughly a few months. The spectrum is harder
when the luminosity is low, and softer and more thermal when the luminosity is
high.

Second-impact emission may be distinguished from intrinsic \ac{AGN} light by
its exceptionally hard spectrum, and the smooth time variation of its
luminosity roughly following the mass-return rate. If the first impact disrupts
the inner disk's corona \citep{2019ApJ...881..113C, 2020ApJ...898L...1R}, then
the few-\si{\kilo\electronvolt} coronal emission of the normal \ac{AGN} would
be reduced, and the tens-of-\si{\kilo\electronvolt} second-impact radiation
would stand out even more.

A third source of radiation not discussed here arises from the unbound debris
interacting with the circumnuclear environment; such debris can be unbound
either during the disruption \citep{2016ApJ...827..127K, 2019MNRAS.487.4083Y},
or by a boost from the disk at the first impact \citep{2019ApJ...881..113C}.
Synchrotron radiation, similar to that of supernova remnants, would be produced
at the shocks generated by the outflowing debris. Such radio signatures have
been observed in \acp{TDE} such as \textacsize{ASASSN}\nobreakdash-14li
\citep{2016ApJ...819L..25A, 2016Sci...351...62V, 2018MNRAS.475.4011B}.
\Acp{TDE} are expected to produce stronger radio emission in \acp{AGN} because
of the higher circumnuclear densities.

Finally, by severely disturbing the disk, \iac{TDE} can leave an enduring mark
on \iac{AGN}. The \ac{AGN} accretion rate may be enhanced for years to come
because half a star has been deposited in the disk. Alternatively, the
accretion rate may be reduced after the end of the \ac{TDE} because the first
impact has emptied the inner disk into the black hole.

To summarize, \acp{TDE} in \acp{AGN} are expected to show a bolometric
luminosity plateau extending for tens of days. This is followed by
\SI{\sim10}{\kilo\electronvolt} to \SI{1}{\mega\electronvolt} hard \xrays{} and
soft \gammarays{} that track the mass-return rate; the typical photon energy
is much harder than in non-jetted \acp{TDE}, and the observed luminosity is
several orders of magnitude below jetted \acp{TDE}. At the same time, the
debris unbound during the disruption and at the first impact can interact with
the circumnuclear gas and produce synchrotron radiation much brighter than
\acp{TDE} in vacuum.

\begin{acknowledgments}
CHC and TP were partially supported by ERC advanced grant \textquote{TReX}. JHK
was partially supported by NSF grant AST-1715032.
\ifapj\par\fi
\end{acknowledgments}

\software{Athena++ \citep{2020ApJS..249....4S}, NumPy
\citep{2020Natur.585..357H}, Matplotlib \citep{2007CSE.....9...90H}}

\ifapj\bibliography{tde}\fi
\ifboolexpr{bool{arxiv} or bool{local}}{\bibhang1.25em\printbibliography}{}

\end{document}